\documentclass[aps,prl,twocolumn,showpacs,nofootinbib,superscriptaddress]{revtex4}
\pdfoutput=1

\usepackage{graphicx}
\usepackage{bm}
\usepackage{amssymb,amsmath,latexsym}
\usepackage{color}
\usepackage{physymb}
\usepackage[colorlinks=true,linktocpage=true,linkcolor=blue,citecolor=blue]{hyperref}

\newcommand{\be}{\begin{equation}}
\newcommand{\ee}{\end{equation}}
\newcommand{\bea}{\begin{eqnarray}}
\newcommand{\eea}{\end{eqnarray}}
\newcommand{\eq}{{\,=\,}}
\newcommand{\Beta}{\bar{\eta}}

\begin{document}


\title{A new exact solution of the relativistic Boltzmann equation\\ and its hydrodynamic limit}

\author{Gabriel S.~Denicol}
\affiliation{Department of Physics, McGill University,
3600 University Street, Montreal, Quebec, H3A 2T8, Canada}
\author{Ulrich Heinz}
\affiliation{Department of Physics, The Ohio State University, Columbus, Ohio 43210-1117, United States}
\author{Mauricio Martinez}
\affiliation{Department of Physics, The Ohio State University, Columbus, Ohio 43210-1117, United States}
\author{Jorge Noronha}
\affiliation{Instituto de F\'{\i}sica, Universidade de S\~{a}o Paulo, S\~{a}o Paulo, SP, Brazil}
\author{Michael Strickland}
\affiliation{Department of Physics, Kent State University, Kent, OH 44242 United States}

\begin{abstract}
We present an exact solution of the relativistic Boltzmann equation for a system undergoing boost-invariant longitudinal and azimuthally symmetric transverse flow (``Gubser flow''). The resulting exact non-equilibrium dynamics is compared to 1st- and 2nd-order relativistic hydrodynamic approximations for various shear viscosity to entropy density ratios. This novel solution can be used to test the validity and accuracy of different hydrodynamic approximations in conditions similar to those generated in relativistic heavy-ion collisions. 
\end{abstract}

\date{\today}

\maketitle



\noindent {\sl 1. Introduction.} Relativistic hydrodynamics plays an important role in the description of the space-time evolution of the quark-gluon plasma (QGP) formed in ultra-relativistic heavy-ion collisions (for a recent review see \cite{Heinz:2013th}). While the ideal limit of the relativistic hydrodynamic equations has been known for a long time~\cite{landau}, the same cannot be stated for relativistic viscous hydrodynamics. The most widespread relativistic formulation of viscous hydrodynamics, relativistic Navier-Stokes theory \cite{landau}, is acausal and intrinsically unstable \cite{acausal}. Its causal generalization remains under intense investigation \cite{degroot,Koide:2006ef,Muronga:2006zx,Baier:2007ix,Bhattacharyya:2008jc,Betz:2008me,PeraltaRamos:2009kg,El:2009vj,Denicol:2010xn,Martinez:2010sc,Florkowski:2010cf,Denicol:2011fa,Denicol:2012cn,Jaiswal:2013vta,Tinti:2013vba,Denicol:2014vaa,Hatta:2014gqa,Hatta:2014gga,Nopoush:2014pfa,Denicol:2014mca,Jaiswal:2014isa}.

For weakly interacting dilute relativistic gases the different approximation schemes that have been used to derive viscous fluid dynamics from first principles can be investigated using the Boltzmann equation as the underlying microscopic theory. To determine which type of hydrodynamic approximation best describes a rapidly expanding QGP, it is useful to consider exactly solvable kinetic models where the approach to thermalization and subsequent hydrodynamic behavior can be studied directly without further approximation. Even though the strongly coupled QGP does not admit a microscopic description via the Boltzmann equation, the macroscopic hydrodynamic description remains valid in the strong-coupling regime as long as the Knudsen number ({\it i.e.} the ratio of mean free path to macroscopic system size) remains small. Hydrodynamic approximations that accurately describe some exactly known microscopic Boltzmann dynamics at weak coupling can therefore be assumed to also provide the most accurate macroscopic description for a strongly coupled liquid. The approach of testing macroscopic hydrodynamic descriptions against exact solutions of the Boltzmann equation has been successfully applied to understand pressure isotropization and thermalization in transversally homogeneous systems undergoing longitudinally boost-invariant expansion \cite{Bjorken:1982qr}, for both initially isotropic \cite{Baym:1984np} and highly anisotropic \cite{Florkowski:2013lza,Florkowski:2013lya,Florkowski:2014sfa} local momentum distributions. So far, none of the available exact relativistic solutions of the Boltzmann equation include simultaneously the effects of longitudinal and transverse expansion.

Transverse expansion of the QGP is, however, the underlying physical mechanism responsible for the collective flow signals observed in ultra-relativistic heavy-ion collisions and used to probe the transport properties of this novel state of nuclear matter. Therefore, to better assess the accuracy of hydrodynamic approximations applied to QGP dynamics, they must be tested under conditions involving simultaneous  transverse and longitudinal expansion. In this Letter we present the first exact solution of the relativistic Boltzmann equation that satisfies this criterium. Our solution is for an azimuthally-symmetric radially expanding boost-invariant conformal system that is undergoing Gubser flow \cite{Gubser:2010ze}. The non-equilibrium dynamics obtained from this exact solution is compared to results obtained using the most popular relativistic hydrodynamic approximations subject to the same flow and symmetries. Specifically, we compare to 1st-order Navier-Stokes (NS) theory \cite{Gubser:2010ze,Gubser:2010ui} and 2nd-order Israel-Stewart (IS) theory \cite{Marrochio:2013wla}. A more detailed derivation of our solution and comparisons to other 2nd-order hydrodynamic approximation schemes can be found in \cite{longpaper}.

\begin{figure}[t]
\includegraphics[width=0.8\linewidth]{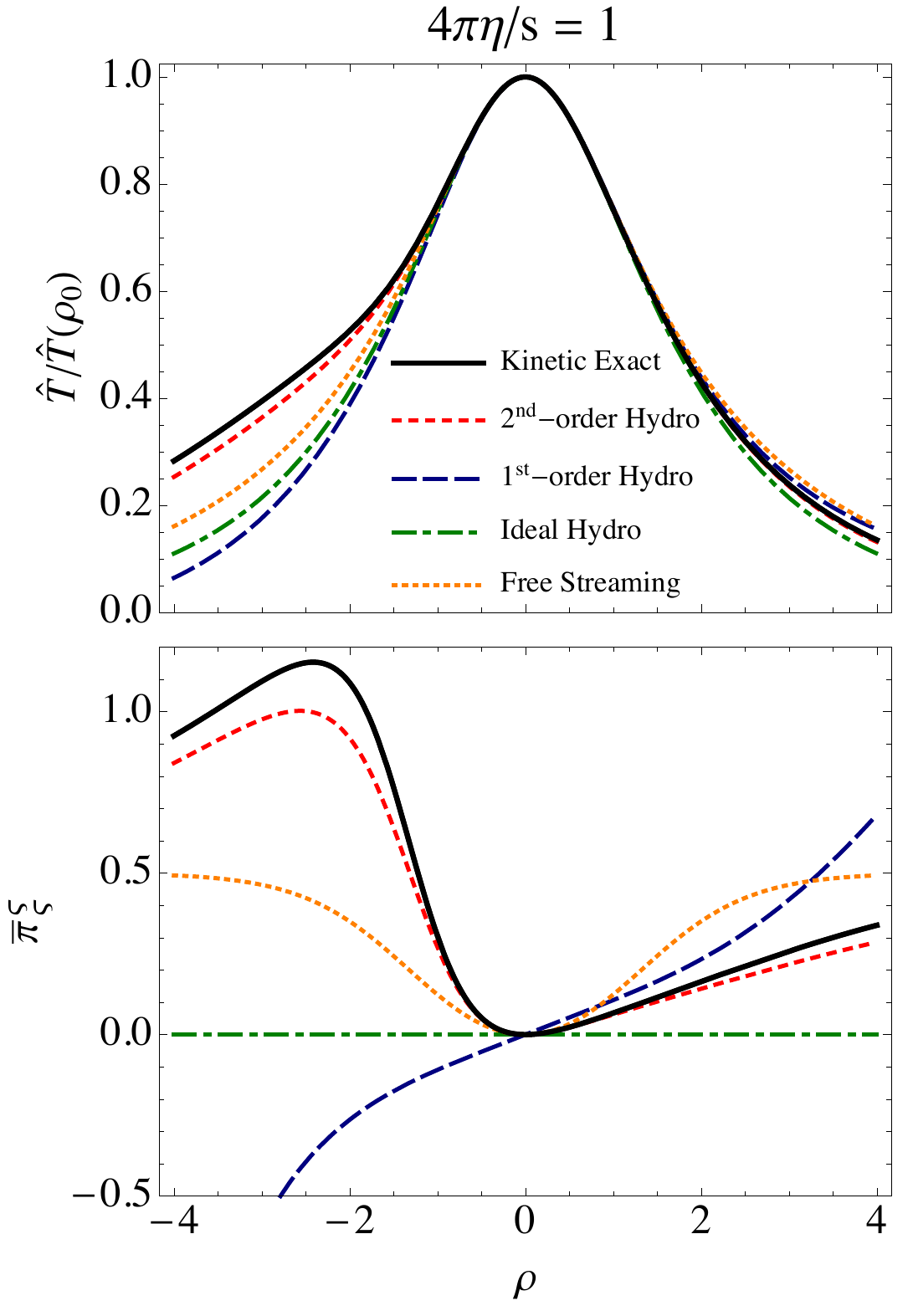}
\vspace{-2mm}
\caption{(Color online) The normalized temperature (top) and shear stress $\bar \pi^\varsigma_\varsigma\equiv \hat \pi_\varsigma^\varsigma/(\hat s\hat T)$ (bottom) as a function of $\rho$ in $dS_3 \otimes R$, obtained from the exact kinetic solution, 2nd-order (IS) and 1st-order (NS) viscous hydrodynamics (all for $\Beta\eq1/(4\pi)$), ideal hydrodynamics ($\Beta\eq0$), and free streaming
($c\eq0$ or $\Beta\eq\infty$). In all cases $\rho_0\eq0$ and $\hat\varepsilon(\rho_0)\eq1$.}
\label{fig:1}
\end{figure}

\begin{figure}[t]
\includegraphics[width=0.8\linewidth]{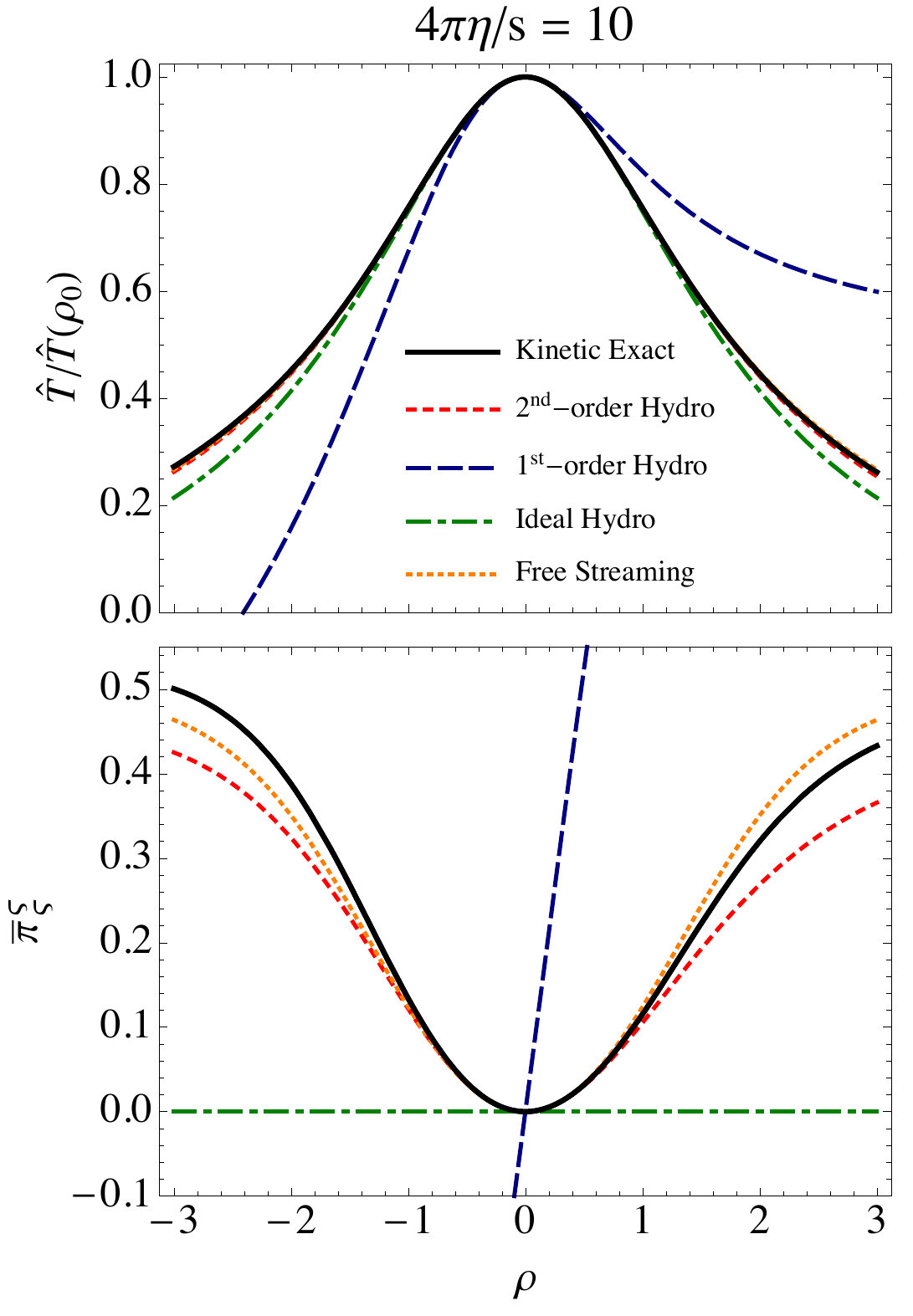}
\vspace{-2mm}
\caption{(Color online) Same as Fig.~\ref{fig:1} except for $\Beta\eq10/(4\pi)$.}
\label{fig:2}
\end{figure}


\noindent {\sl 2. Exact (1+1)-dimensional kinetic solution.} The dynamics of high-energy heavy-ion collisions is usually described in Minkowski space-time using Milne coordinates $x^\mu\eq(\tau,x,y,\varsigma)$, with longitudinal proper time $\tau\eq\sqrt{t^2{-}z^2}$ and space-time rapidity $\varsigma\eq\atanh(z/t)$ and metric $g_{\mu\nu}\eq(-1,1,1,\tau^2)$. In this paper, the fluid velocity $u^\mu$ is taken to be the Gubser flow \cite{Gubser:2010ze} which possesses a conformal $SO(3)_q\otimes SO(1,1)\otimes Z_2$ symmetry (Gubser symmetry \cite{Gubser:2010ze,Gubser:2010ui}). Systems with such a flow profile are more conveniently described in a curved space-time given by the direct product of 3-dimensional de Sitter space and a line, $dS_3\otimes R$ \cite{fn2}. The two space-times are related by a Weyl rescaling of the metric, $d\hat s^2\eq{ds}^2/\tau^2$, with $ds^2\eq-d\tau^2{+}dr^2{+}r^2d\phi^2{+}\tau^2d\varsigma^2$ being the Minkowski line element in polar Milne coordinates, and $d\hat{s}^2\eq-d\rho^2+\cosh^2\rho\left(d\theta^2+\sin^2\theta d\phi^2\right)+d\varsigma^2$ being the line element in $dS_3\otimes R$. Here we introduced the de Sitter coordinates $\hat{x}^\mu\eq(\rho,\theta,\phi,\varsigma)$, with \cite{Gubser:2010ui}
\be
\label{eq:rhotheta}
\begin{split}
\rho(\tau,r) &= -\asinh\left(\frac{1-q^2 \tau^2+q^2 r^2}{2q\tau}\right) , \\
\theta(\tau,r) &= \atan\left(\frac{2 q r}{1+q^2 \tau^2-q^2 r^2}\right) .
\end{split}
\ee
$q^{-1}$ is an arbitrary length scale that sets the size of the system. In these coordinates, the Gubser flow profile $u_\tau\eq{-}\cosh\kappa(\tau,r)$, $u_r\eq\sinh\kappa(\tau,r)$, with transverse flow rapidity $\kappa\eq\atanh\bigl(2q^2\tau r/(1{+}q^2\tau^2{+}q^2r^2)\bigr)$ \cite{Gubser:2010ze,Gubser:2010ui}, simplifies to $\hat u_\mu\eq(-1,0,0,0)$~\cite{Gubser:2010ze,Gubser:2010ui}, {\it i.e.} the system becomes macroscopically static. (All quantities in $dS_3\otimes R$ are denoted by a ``hat''.)

We seek an exact solution of the Boltzmann equation
\be
p^\mu \partial_\mu f(x,p)={\mathcal C}[f](x,p) \, ,
\label{boltzmanneq}
\ee 
with a simplified \cite{fn1} collision term in relaxation time approximation (RTA) \cite{aw,book}
\be
\label{eq:RTA}
\mathcal{C}[f](x,p) = \frac{p{\cdot}u(x)}{\tau_\mathrm{rel}(x)}\bigl(f(x,p){-}f_{\rm eq}(x,p)\bigr).
\ee
Here $f_{\rm eq}(x,p)$ is the local equilibrium distribution function, assumed to be of Boltzmann form $f_{\rm eq}\eq\exp(p{\cdot}u/T)$, where $T(x)$ is the local temperature, related by conformal invariance to the relaxation time by $\tau_{\rm rel}(x)\eq{c}/T(x)$. $c$ is a free dimensionless parameter that, in RTA, can be expressed in terms of the shear viscosity to entropy density ratio $\Beta\equiv\eta/s$ as $c=5\Beta$ \cite{Denicol:2010xn,Denicol:2011fa,Florkowski:2013lza,Florkowski:2013lya}. 

The Gubser symmetry severely restricts the possible combinations of the coordinates $\hat x^\mu$ and momenta $\hat p^\mu$ on which the distribution function $f(\hat{x},\hat{p})$ can depend. First, conformal invariance requires massless degrees of freedom, {\it i.e.} $\hat{p}^2\eq0$, so $\hat{p}_\rho$ can be eliminated in terms of the ``spatial'' momentum components: $\hat{p}_\rho\eq\sqrt{(\hat{p}_\theta/\cosh\rho)^2{+}(\hat{p}_\phi/(\cosh\rho\sin\theta))^2{+}\hat{p}_\varsigma^2}$. $Z_2$ symmetry requires $f$ to be an even function of $\varsigma$. The $SO(1,1)$ symmetry implements longitudinal boost-invariance which reduces the dependence on the longitu\-dinal coordinates and momenta to the combination $\hat{p}_\varsigma\eq\tau p_T\sinh(y{-}\varsigma) \eq{t}p_z{-}zE$ (where $p_T\eq\sqrt{p_x^2{+}p_y^2}$ is the transverse momentum and $y$ is the momentum rapidity). Symmetry under the $SO(3)_q$ subgroup implies that the distribution function is independent of $\theta$ and $\phi\eq\atan(y/x)$ and that it depends only on the following combination of momentum components \cite{longpaper}:
\be
\label{eq:momrho}
\hat{p}_\Omega^2=\hat{p}_\theta^2+\frac{\hat{p}_\phi^2}{\sin^2\theta}\,.
\ee
In other words, $f(\hat{x};\hat{p})\eq{f}(\rho;\hat{p}_\Omega^2,\hat{p}_\varsigma)$ for Gubser-symmetric systems. 

\begin{figure*}
\begin{center}
\includegraphics[width=0.92\linewidth]{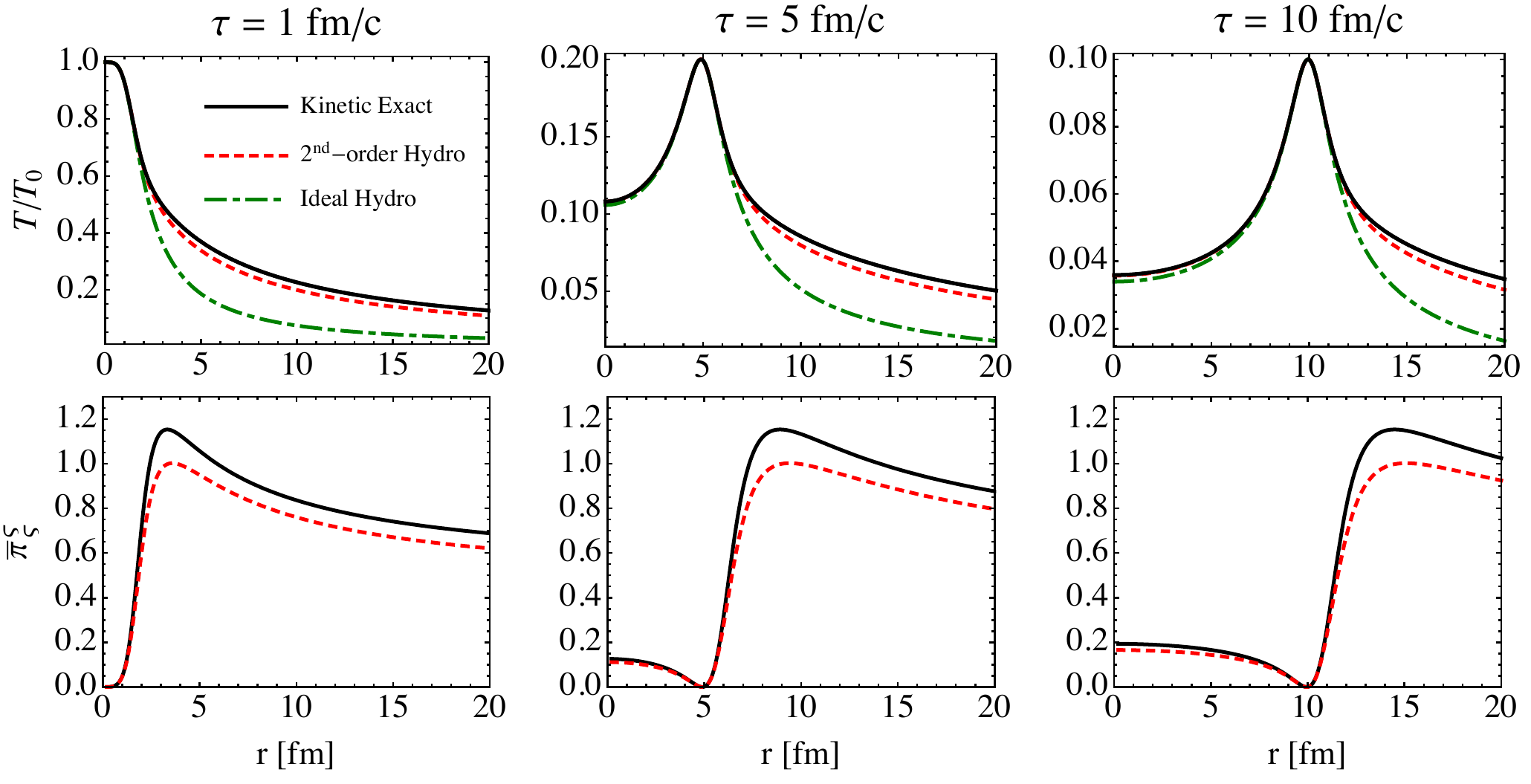}
\end{center}
\vspace{-5mm}
\caption{(Color online) Snapshots of the radial profiles of the temperature (top) and the shear stress component $\bar{\pi}^\varsigma_\varsigma$ (bottom) computed at $\tau=1$, 5, and 10 fm/$c$ (with $q\eq1$\,fm$^{-1}$). The results shown correspond to the exact kinetic solution (solid black line), ideal hydrodynamics (dot-dashed green line), and 2nd-order IS theory (red short-dashed line). In this figure 
$\Beta\eq1/(4\pi)$.}
\label{fig:3}
\end{figure*}

Using these constraints, the RTA Boltzmann equation in de Sitter coordinates reduces \cite{longpaper} to the following simple relaxation-type equation:
\be
\frac{\partial}{\partial \rho}f(\rho;\hat p_\Omega^2,\hat p_\varsigma)= -\frac{\hat{T}(\rho)}{c}\Bigl[f(\rho;\hat p_\Omega^2,\hat p_\varsigma)-f_{\rm eq}\Bigl(\frac{\hat p^\rho}{\hat T(\rho)}\Bigr)\Bigr].
\label{newRTAboltzmanneq}
\ee
Here $\hat p^\rho\eq\sqrt{(\hat p_\Omega/\cosh\rho)^2{+}\hat p_\varsigma^2}$ and $\hat{T}\eq\tau T$. As had been shown in \cite{Gubser:2010ui} for the hydrodynamic case, imposing Gubser symmetry reduces the partial differential equation (\ref{boltzmanneq}) to a simple ordinary differential equation \cite{comment1}. 

Equation~\eqref{newRTAboltzmanneq} has the following exact solution \cite{Florkowski:2013lza,Florkowski:2013lya}
\be
\begin{split}
f(\rho;\hat p_\Omega^2,\hat p_\varsigma) &=D(\rho,\rho_0) f_0(\rho_0;\hat p_\Omega^2,\hat p_\varsigma)\,,\\
&\hspace{-0.7cm}+\frac{1}{c}\int_{\rho_0}^\rho d\rho'\,D(\rho,\rho')\,\hat T(\rho')\, f_{\rm eq}(\rho';\hat p_\Omega^2,\hat p_\varsigma) \, ,
\label{boltzmannsolution}
\end{split}
\ee
where $D(\rho,\rho_0)=\exp\!\left[-\int_{\rho_0}^\rho d\rho' \,\hat T(\rho')/c\right]$ is the damping function and  $f_0(\rho_0;\hat p_\Omega^2,\hat p_\varsigma)$ is the initial distribution function at $\rho_0$. In the following we assume that the initial configuration corresponds to an equilibrated state; other possible choices will be studied in future work. 

All macroscopic hydrodynamic quantities can now be computed from their kinetic definitions as momentum moments of $f$. For example, the energy density is given by $\hat \varepsilon (\rho)\eq\int\,d\hat{P}\,(\hat{p}^\rho)^2\,f(\rho;\hat p_\Omega^2,\hat p_\varsigma)$ and the shear stress tensor by $\hat \pi_{\mu\nu}\eq\int\,d\hat{P}\,\hat \Delta^{\alpha\beta}_{\mu\nu}\,\hat p_\alpha \hat p_\beta\,f(\rho;\hat p_\Omega^2,\hat p_\varsigma)$, with $d\hat{P}\eq{d}\hat p_\varsigma\,d\hat p_\theta\,d\hat p_\phi/\left[(2\pi)^3\,\hat p^\rho\cosh^2\rho\,\sin\theta\right]$. $\hat \Delta^{\mu\nu}_{\alpha\beta}$ is the transverse and traceless double projector in $(\rho,\theta,\phi,\varsigma)$ coordinates. By taking momentum moments of \eqref{boltzmannsolution}, exact integral equations similar to (\ref{boltzmannsolution}) can be written for $\hat \varepsilon$ and $\hat \pi_{\mu\nu}$ \cite{longpaper}. These equations are solved numerically by the iterative procedure described in Ref.~\cite{Florkowski:2013lya}. The resulting energy density defines the temperature and, hence, the equilibrium distribution function for $\rho\in (-\infty,\infty)$. Once this is done, the full non-equilibrium distribution function can also be determined from Eq.~\eqref{boltzmannsolution}. 

\begin{figure*}
\begin{center}
\includegraphics[width=0.45\linewidth]{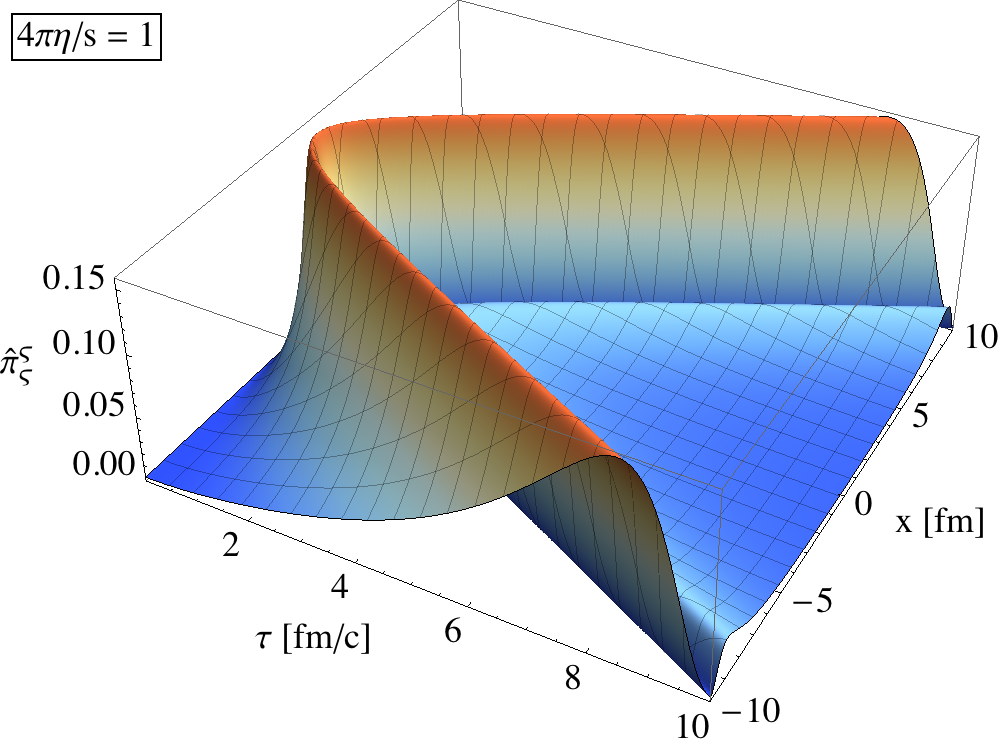} 
$\;\;\;$
\includegraphics[width=0.45\linewidth]{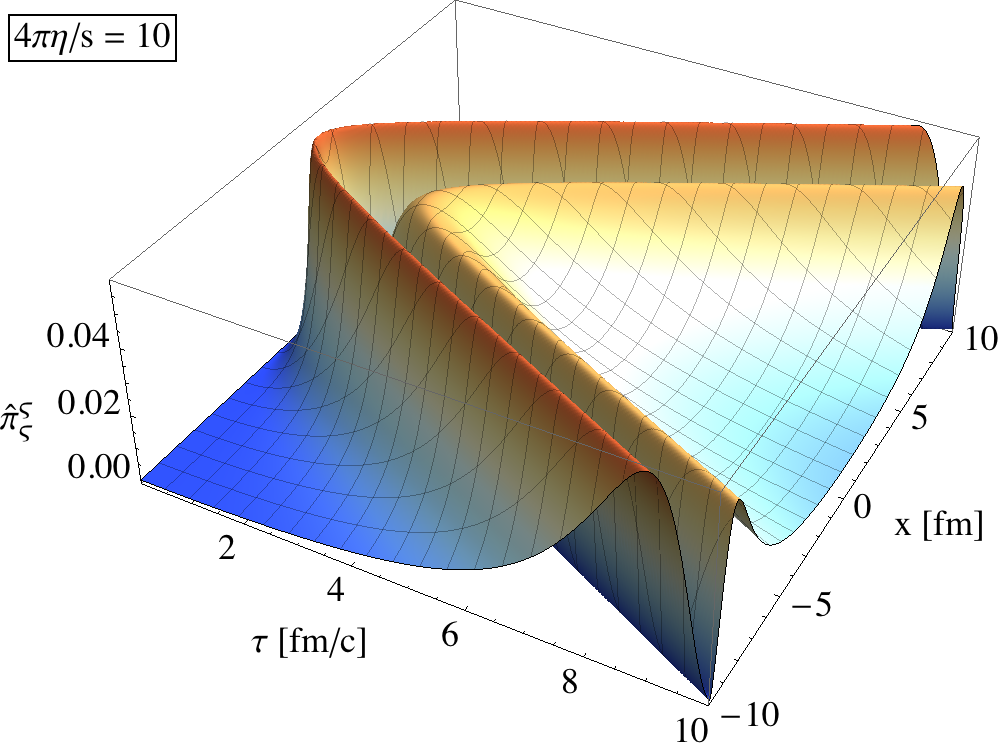}
\end{center}
\caption{(Color online) Two-dimensional slice showing, for scale parameter $q\eq$1\,fm$^{-1}$, the Minkowski space-time evolution at $y\eq{z}\eq0$ of the shear stress $\hat{\pi}^{\varsigma\varsigma}= \tau^4 \pi^\varsigma_\varsigma$, for $\Beta\eq1/(4\pi)$ (left panel) and $\Beta\eq10/(4\pi)$ (right panel).}
\label{fig:4}
\end{figure*}

In this Letter we compare the evolution of energy density, temperature, and shear stress obtained from the above exact kinetic solution to previously found solutions corresponding to three different hydrodynamic approximations: ideal hydrodynamics ($c\eq0$), NS theory \cite{Gubser:2010ze}, and conformal IS theory \cite{Marrochio:2013wla}. We also compare with the analytic free streaming limit which can be obtained from the kinetic theory solution (\ref{newRTAboltzmanneq}) by setting $c\to\infty$. The ideal fluid solution is \cite{Gubser:2010ze,Gubser:2010ui}
\be
\label{eq:ideal}
\hat\varepsilon(\rho)=\frac{\hat\varepsilon(\rho_0)}{\left(\cosh\rho\right)^{8/3}}\,,
\qquad
\hat\pi^{\mu\nu} (\rho)=0\,. 
\ee
Viscous hydrodynamics with Gubser flow is fully characterized by a single shear stress component \cite{Marrochio:2013wla} for which we will choose $\bar \pi^\varsigma_\varsigma\equiv \hat \pi^\varsigma_\varsigma/(\hat{s}\hat{T})$ (where $\hat{s}\eq\tau^3 s$ is the entropy density in de Sitter space). The IS equation for the temperature obtained from the conservation of energy-momentum, $\hat\nabla_\mu\hat T^{\mu\nu}=0$, is
\be
\label{eq:visceqtemp}
\frac{1}{\hat T}\frac{d\hat T}{d\rho} +\frac{2}{3}\tanh\rho = \frac{\bar \pi^\varsigma_\varsigma}{3}\,\tanh \rho\,,
\ee
while for the shear stress component one finds
\be
\label{eq:ISeqpietaeta}
\frac{d\bar \pi^\varsigma_\varsigma}{d\rho} 
+ \frac{\hat T}{c\Beta}\bar\pi^\varsigma_\varsigma
+ \frac{4}{3}\Bigl[\left(\bar \pi^\varsigma_\varsigma\right)^2{-}\frac{1}{c}\Bigr]\,\tanh\rho 
= 0\,.
\ee
In Eq.~\eqref{eq:ISeqpietaeta} we use the same relaxation time as in Eq.~\eqref{eq:RTA}, namely $\tau_{\rm rel} = 5 \bar\eta/T$. The NS result of \cite{Gubser:2010ze,Gubser:2010ui} corresponds to the $c \to 0$ limit of Eq.~(\ref{eq:ISeqpietaeta}) which gives $(\bar\pi^\varsigma_\varsigma)_{\rm NS} =\frac{4}{3}\frac{\Beta}{\hat T}\tanh\rho$. 

The $dS_3\otimes R$ solutions for $\hat \varepsilon$ and $\hat{\pi}^{\mu\nu}$  can be translated back to Minkowski space via the prescription \cite{Gubser:2010ze,Gubser:2010ui,Marrochio:2013wla}
\be
\label{eq:dictionary}
\begin{split}
{\varepsilon}(\tau,r)&=\frac{\hat\varepsilon\left(\rho(\tau,r)\right)}{\tau^4}\,,\\
\pi_{\mu\nu}(\tau,r)&=\frac{1}{\tau^2}\frac{\partial\hat x^\alpha}{\partial x^\mu}\frac{\partial\hat x^\beta}{\partial x^\nu}\hat\pi_{\alpha\beta}\left(\rho(\tau,r)\right)\,.
\end{split}
\ee
%


\noindent {\sl 3. Numerical results.} In Figures~\ref{fig:1} and \ref{fig:2} we show the de Sitter time evolution of the normalized temperature $\hat T/\hat T(\rho_0)$ (top panel) and shear stress $\bar \pi^\varsigma_\varsigma$ (bottom panel), with initial conditions $\hat\varepsilon(\rho_0)=1$ and $\bar\pi^{\varsigma}_{\varsigma}(\rho_0)\eq0$ at $\rho_0\eq0$, for the exact kinetic solution, its free streaming limit, and three different hydrodynamic approximations: ideal as well as NS and IS viscous hydrodynamics. For small specific shear viscosity $\Beta\eq1/(4\pi)$, shown in Fig.~\ref{fig:1}, 2nd-order IS viscous fluid dynamics provides the best description for both quantities; for ten times larger shear viscosity $\Beta\eq10/(4\pi)$, shown in Fig.~\ref{fig:2}, the free-streaming limit provides the best approximation to the exact solution, again followed by 2nd-order IS theory as the best hydrodynamic approximation. The improvement resulting from including 2nd-order corrections is particularly evident for $\bar\pi^{\varsigma}_{\varsigma}$:
1st-order NS theory exhibits qualitatively wrong behavior for $\bar\pi^{\varsigma}_{\varsigma}$ at negative $\rho$ values. 

Figure~\ref{fig:3} shows snapshots of the radial temperature (top) and shear stress (bottom) profiles in Minkowski space at three longitudinal proper times, $\tau=1$, 5, and 10 fm/$c$, choosing $q\eq1$\,fm$^{-1}$ for the scale parameter. For $\Beta\eq1/(4\pi)$, the exact kinetic solution is compared with ideal and 2nd-order viscous IS hydrodynamics. Once again, 2nd-order IS theory is found to provide a good description of the kinetic results. One may observe that at no fixed value for the longitudinal proper time the different approximations agree with each other perfectly. Although they are equal at the initial de Sitter time $\rho_0$, where we assumed the system to be in local equilibrium, Eq.~\eqref{eq:rhotheta} shows that surfaces of constant $\rho$ translate into hyperbola-like surfaces in the $\tau$-$r$ plane, and that at fixed $\tau$, small (large) $r$ values correspond positive (negative) values of $\rho{-}\rho_0$ where the system is out of equilibrium. Even for $\Beta$ as small as $1/(4\pi)$, ideal hydrodynamics is seen to fail badly at large $r$ values ({\it i.e.} at large negative $\rho$) whereas 2nd-order hydrodynamics performs much better. This is consistent with Figs.~\ref{fig:1} and \ref{fig:2}. The location in Fig.~\ref{fig:3} of the peak value of the temperature corresponds to $\rho\eq\rho_0$, and at this point all approximations agree by construction.

Finally, we present in Fig.~\ref{fig:4} a 3d visualization of the full space-time evolution of the exact kinetic theory result for the shear stress $\hat{\pi}^\varsigma_\varsigma = \tau^4 \pi^\varsigma_\varsigma$, for
$4\pi\Beta=1$ and 10 in the left and right panels, respectively. Clearly, the shear stress is strongly affected by an increase of the shear viscosity by a factor 10. The location of the trough corresponding to $\hat{\pi}^\varsigma_\varsigma\eq0$ indicates the initial condition surface $\rho_0\eq0$, and the inner region bounded by it represents positive $\rho$ values. 


\noindent {\sl 4. Conclusions.} We presented an analytic (1+1)-d solution of the RTA Boltzmann equation for a system with Gubser flow. This is the first exact solution of the Boltzmann equation for a relativistic system undergoing simultaneous longitudinal and transverse expansion. It allows one to compute all components of the energy-momentum tensor and to study their full space-time evolution exactly. By comparing these exact solutions with the most widespread formulations of relativistic hydrodynamics we were able to determine their regime of validity at each space-time point of for varying values of the specific shear viscosity $\Beta$.
We found that, quite generally, 2nd-order viscous hydrodynamics provides an overall good description of the exact kinetic results and performs much better than relativistic Navier-Stokes theory. The exact solution obtained here opens novel ways to test the accuracy of different hydrodynamic approaches used to describe the dynamics of the QGP formed in ultra-relativistic heavy-ion collisions. Future analysis of the solution \eqref{boltzmannsolution} in momentum space with non-equilibrium initial conditions promises to offer detailed insights into the dynamics of longitudinal/transverse momentum isotropization and thermalization in relativistic systems undergoing simultaneously transverse and longitudinal expansion.


\noindent {\sl Acknowledgments:} G.S.D. was supported by a Banting Fellowship from the Natural Sciences and Engineering Research Council of Canada.  U.H. and M.M. acknowledge support from the U.S. Department of Energy, Office of Science, Office of Nuclear Physics under Award No.~\rm{DE-SC0004286}. J.N. thanks the Conselho Nacional de Desenvolvimento Cient\'ifico e Tecnol\'ogico (CNPq) and Funda\c c\~ao de Amparo \`a Pesquisa do Estado de S\~ao Paulo (FAPESP) for support. U.H.\ and M.S.\ were supported in part (in the framework of the JET Collaboration) by U.S. DOE Awards No.~\rm{DE-SC0004104} and \rm{DE-AC0205CH11231}. Finally, U.H., M.M. and J.N. acknowledge support through a bilateral scientific exchange program between FAPESP and the Office of Sponsored Programs at The Ohio State University.



\begin{thebibliography}{99}
\bibitem{fn2} We emphasize that this is a convenient transformation that facilitates the implementation of the symmetries necessary to solve the Boltzmann equation. The solution in flat space is obtained by mapping our solution from de Sitter coordinates to Milne coordinates, where the direct implementation of Gubser flow and Gubser symmetric boundary conditions would have been much more difficult.  By construction, the resulting solution in Milne coordinates solves the Boltzmann equation in these coordinates subject to the Gubser flow and boundary conditions.
\bibitem{fn1}
Other (linearized) forms for the collision kernel involving multiple microscopic time scales can in principle be implemented in the following approach but will not be considered here.
\bibitem{comment1} Note that the same reduction to an ordinary differential equation still happens if the RTA collision term (\ref{eq:RTA}) is replaced by the full non-linear (in $f$) Boltzmann collision integral $C[f](\rho;\hat p_\Omega^2,\hat p_\varsigma)$.
\end{thebibliography}
\end{document}